\documentclass[10pt,journal,compsoc]{IEEEtran}
\ifCLASSOPTIONcompsoc
 \usepackage{lineno,hyperref}
 \usepackage[english]{babel}
  \usepackage[utf8]{inputenc}
  \usepackage{amsmath}
  \usepackage{amssymb}
  \usepackage{graphicx}
  \usepackage{indentfirst}
  \usepackage{nicefrac}\DeclareUnicodeCharacter{00A0}{ }
  \usepackage[colorinlistoftodos]{todonotes}
\usepackage{algorithmic,algorithm}
\usepackage{multirow}
\usepackage{xcolor}
 \usepackage{amsmath,booktabs}
  \usepackage{graphicx}
  \usepackage{epstopdf}
  \usepackage{epsfig}
  \usepackage{multirow}
  \usepackage{float}
  \usepackage{color}
  \usepackage{bbding}
  \usepackage{color, colortbl}
  \usepackage{longtable,lscape}
  \usepackage{color, colortbl}
  \usepackage{bigstrut}
  \usepackage{pifont}
 \usepackage{ragged2e}
  \usepackage{wasysym}
  \usepackage{amssymb}
  \usepackage{adjustbox}
  \usepackage{caption}
\usepackage{multirow}
\usepackage{mathtools}
\usepackage{caption}
\usepackage{subcaption}
  \usepackage{indentfirst}
  \usepackage{adjustbox}

\modulolinenumbers[5]
 
\usepackage{tabularx}

\newcounter{protocol}

  \usepackage[nocompress]{cite}
\else
  \usepackage{cite}
\fi
\ifCLASSINFOpdf
\else
\fi
\begin{document}
%
\title{Modification to Fully Homomorphic Modified Rivest Scheme }
\author{
		Sona~Alex~ and~
	Bian~Yang
	\thanks{The authors are with the Department of Information Security and Communication Technology at the Norwegian University of Science and Technology (NTNU), Gjøvik, Norway. (E-mail: sona.alex@ntnu.no; bian.yang@ntnu.no)}}
\markboth{}%
{Shell \MakeLowercase{\textit{et al.}}: Bare Demo of IEEEtran.cls for IEEE Journals}
\maketitle
\IEEEdisplaynontitleabstractindextext
%
\IEEEpeerreviewmaketitle
This document details the Fully Homomorphic Modified Rivest Scheme (FHMRS) \cite{gfc}, security issue in FHMRS  and modification to FHMRS (mFHMRS) to mitigate the security issue.
The following section details FHMRS.
\section{ Fully Homomorphic Modified Rivest Scheme (FHMRS)}
This section details the symmetric-key-based fully homomorphic modified  Rivest scheme (FHMRS) presented in \cite{gfc}.
The following are the various functions of FHMRS:
\begin{itemize}
\item \textbf{KeyGen($\lambda$, $\mathbb{M}$, $N$ ):} Accepts as inputs security parameter ($\lambda$) that decides the security in terms of the computational complexity of attackers, the message space $\mathbb{M}$ and number of consecutive multiplications $N$ supported by the scheme. Message space is selected so that the results after computations on the input messages should come within the message space. Based on the computations on the input messages, the size of the output message (computational results)  will be higher than the size of the input messages. For example, assume $l_m$ is the size of input messages and that there are N consecutive multiplications and A consecutive additions. Then $log_{2} (\mathbb{M}) > (N+1)* l_m + A$. Generate three primes $p$, $q$ and $u$ where $p$ and $q$ are equal in size, $\mathbb{M} < u << (p* q=n) $. $p$, $q$ and $u$ are the secret parameters of FHMRS.\\
\item \textbf{Encrypt($m_{i}, p, q, u$):} Accepts secret primes ($p, q$ and $u$) and the $i^{th}$ message $m_{i}$ $\in$ $\mathbb{Z}_{\mathbb{M}}$ to be encrypted as inputs. Choose a random number $0<$$g_{i}$ $<u$. The ciphertext $c_{i}=(c_{i1},c_{i2})$ is computed relying on the Chinese Remainder Theorem (CRT) as shares of $m_{i}$ as follows.
\begin{equation}
\label{eqn:e4}
c_{i}=([(m_{i}+(g_{i}* u))]_{p}, [(m_{i}+(g_{i}* u)]_{q})=(c_{i1},c_{i2})
\end{equation}
Here,  $m_{i}+(g_{i}* u) < (p* q)$ to recover it correctly based on the Chinese remainder theorem during decryption.
where $[]_{p}$ indicates modulo p operation. $[(m_{i}+(g_{i}* u))]_{p}$ is notated as $FHMRS_{p}(m_{i})$ from hereafter.
\item \textbf{ADD($c_{1},c_{2}$):} The two ciphertexts $c_{1}$ $=$ $(FHMRS_{p}(m_{1})$, $FHMRS_{q}(m_{1}))$$= (c_{11},c_{12})$ and $c_{2} =(FHMRS_{p}(m_{2}),$ $FHMRS_{q}(m_{2})) = (c_{21},c_{22})$ are inputs to ADD function. ADD executes $(c_{1}+_{a}c_{2})$ as $(c_{11}+c_{21})$ and $(c_{12}+ c_{22})$ and generates the ciphertext ($FHMRS_{p}(m_{1}+m_{2})$, $FHMRS_{q}(m_{1}+m_{2})$). $+_{a} (-_{ a} )$ represents homomorphic addition (subtraction) of FHMRS scheme.
\item \textbf{ConstADD($c_{i}, t$):} A constant $t$ and a ciphertext $c_{i} =$$(FHMRS_{p}(m_{i})$, $FHMRS_{q}(m_{i}))$ $= (c_{i1},c_{i2})$ are the inputs to ConstADD function. ConstADD executes $(t +_{ca} c_{i})$ as $( (t+c_{i1}), (t+c_{i2}))$ and generates the ciphertext $c_{cadd}$= ($FHMRS_{p}(t+m_{i})$, $FHMRS_{q}(t+m_{i})$). $+_{ca} (-_{ca})$
represents homomorphic constant addition (subtraction) in FHMRS.
\item \textbf{MUL($c_{1},c_{2}$):} The two ciphertexts $c_{1}$ $=$ $(FHMRS_{p}(m_{1})$, $FHMRS_{q}(m_{1}))= (c_{11},c_{12})$ and $c_{2} =(FHMRS_{p}(m_{2}),$ $FHMRS_{q}(m_{2}))= (c_{21},c_{22})$ are inputs to MUL function. MUL executes $(c_{1}*_{m}c_{2})$ as $(c_{11}*c_{21})$ and $(c_{12}* c_{22})$ and generates the ciphertext $c_{mul}$ = ($FHMRS_{p}(m_{1}*m_{2})$, $FHMRS_{q}(m_{1}*m_{2})$)=$ ([(m_{1}+(g_{1}* u))]_{p}*[(m_{2}+(g_{2}* u))]_{p}, [(m_{1}+(g_{1}* u))]_{q}*[(m_{2}+(g_{2}* u))]_{q} )$.  ($FHMRS_{p}(m_{1}*m_{2})$, $FHMRS_{q}(m_{1}*m_{2})$) represents the ciphertext after homomorphic multiplication of ciphertexts of $m_{1}$ and $m_{2}$. $*_{m}$ represents the multiplicative homomorphism of the FHMRS scheme.
\item \textbf{ConstMUL($c_{i}, t$):} A constant $t$ and a ciphertext $c_{i} =$ $(FHMRS_{p}(m_{i})$, $FHMRS_{q}(m_{i}))$ $= (c_{i1},c_{i2})$ are inputs to ConstMUL function. ConstMUL executes $(t *_{cm} c_{i})$ as $( (t*c_{i1}), (t*c_{i2}))$ and generates the ciphertext $c_{cmul}$= ($FHMRS_{p}(t*m_{i})$, $FHMRS_{q}(t*m_{i})$). $*_{cm}$
represents homomorphic constant multiplication in the FHMRS scheme.
\item \textbf{Decrypt($c_{i}, p, q, u$) :} This function takes the ciphertext $c_{i} = (c_{i1}, c_{i2})$ corresponding to $i^{th}$ message $m_{i}$ and secret primes ($p$, $q$ and $u$) of FHMRS as inputs. Based on CRT, the message ($m_{i}$) is computed from $c_{i}$ as in Eqn. \ref{eqn:e5}.
\begin{equation}
\label{eqn:e5}
m_{i} =\big[ \big[([c_{i1}]_{p} * q* [q^{-1}]_{p}) +( [c_{i2}]_{q}* p* [p^{-1}]_{q})\big ]_{ n}\big]_{u} .\\
\end{equation}
Where, $ \big[([c_{i1}]_{p}* q * [q^{-1}]_{p}) +( [c_{i2}]_{q}* p* [p^{-1}]_{q})\big ]_{ n}=CRT([c_{i1}]_{p}, [c_{i2}]_{q})$ \\
\end{itemize}
Here, $\mathbb{M} < u$ so that modular reduction by $u$ in the decryption function will not alter the message. During decryption, the terms involving u in the ciphertext will be cancelled.  \\
Note: If there are N consecutive multiplications, $l_{n} > \big(((N+1)*(l_{u}+l_{g})) + (N+1)\big)$. Here, $l_{n}$, $l_{u}$ and $l_{g}$ indicate the number of bits required to represent $n=p* q$, $u$ and $g$, respectively. The size of the resulting product after multiplications is decided by the size of the biggest term and the increase in the bit-length due to additions. Here, the biggest term in the multiplicative ciphertext after N successive ciphertext multiplications would be the product of (N+1) number of random numbers ($g_{1}, g_{2},...,  g_{N+1}$) and $u^{N+1}$ since the bit-length of $m_{i}$, $l_{m_{i}} < \frac{l_{u}}{N}$. Hence, the size of the largest term in the product is $\big((N+1)*(l_{u}+l_{g})$. Even though N consecutive ciphertext multiplication involves $2^{N+1} -1$  additions, it increases the size of the product by a maximum of (N+1) bits since $l_{m_{i}} < \frac{l_{u}}{N}$. Since the size of the resulting product after multiplications should be less than $n$  for the correctness of the decryption function that involves modular $n$ operation based on CRT, $l_{n} > \big((N+1)*(l_{u}+l_{g}) +1)\big)$.\\
If there are N consecutive multiplication and A consecutive additions, then for the correctness of the decryption function that involves modular $n$ operation based on CRT, $l_{n} > \big((N+1)*(l_{u}+l_{g} + 1\big)+ A$.\\
\section{ Known plaintext attack on FHMRS when it supports multiplicative homomorphism}
\label{gkpa}
This section describes the known plaintext attack on FHMRS when it supports multiplicative homomorphism.
In FHMRS, if it supports N consecutive multiplications and A consecutive additions, the size of $n$ in bits,  $l_{n} > \big((N+1)*(l_{u}+l_{g}+1)\big)+ A$ and the size of $p$ and $q$ are $l_{p}=l_q= l_n/2$. In the encryption function, modular reduction on $(m_{i}+(g_{i}* u))$ by $p$ and $q$ is performed. The size of $(m_{i}+(g_{i}* u))$ in bits is $(l_{u}+l_{g}+1)$. When FHMRS supports N multiplication and A additions, $l_{p}=l_q> \dfrac{\big((N+1)*(l_{u}+l_{g}+1)\big)+ A}{2}$. Hence, $l_{p}=l_q > (l_{u}+l_{g}+1)$ and modular reduction on $(m_{i}+(g_{i}* u))$ by $p$ and $q$ results in $(m_{i}+(g_{i}* u))$ itself. Suppose an attacker possesses two ciphertext-plaintext pairs; he can mount KPA. Let us assume attacker has two plaintexts $m_1$ and $m_2$ and corresponding ciphertexts  $c_1 = (m_{1}+(g_{1}* u))$ and  $c_2 =(m_{2}+(g_{2}* u))$, respectively. Then the attacker can find the greatest common divisor (GCD) of $c_1-m_1$ and $c_2-m_2$, since $c_1-m_1=(g_{1}* u) $ and $c_2-m_2=(g_{2}* u)$. The GCD will be the secret $u$. To mitigate this issue, we have modified FHMRS as given in the following section.
\section{Modified FHMRS (mFHMRS)}
This section details the modification to FHMRS to mitigate the known-plaintext attack mentioned in Section 2.
\begin{itemize}
\item \textbf{KeyGen ($\lambda$, $\mathbb{M}$, $N$, $A$):} Accepts as inputs security parameter ($\lambda$) that decides the security in terms of the computational complexity of attackers, the message space $\mathbb{M}$ and number of consecutive multiplications $N$ and consecutive additions $A$ supported by the scheme. If $l_m$ is the size of input messages,  $log_{2} (\mathbb{M}) > (N+1)* l_m + A$. Generate primes $p_1,p_2, ..., p_{N+S}$ and $u$ where $p_1, p_2,...,p_{N+S}$ are equal in size, $\mathbb{M}<u < (p_1* p_2*...* p_{N+S}=n) $.  Let $u$ and $l_{p_{i}}$ be $l_u$-bit and $l_p$-bit primes, respectively. Also, $p_1,p_2, ..., p_{N+S}$ should be large enough to prevent guessing $p_i$s from the range of values of corresponding ciphertext share. $l_p=l_{p_1}= l_{p_2}= \dots = l_{p_{(N+S)}} \ge \dfrac{(N+1)}{(N+S)}*(l_u +l_g+1)+\dfrac{A}{(N+S)}$, where $l_g$ is the number of bits in the random number used in the encryption function. $N+S$ is the number of components or shares in the ciphertext, where $S$ is selected such that $\dfrac{(N+1)}{(N+S)}*(l_u +l_g+1)+\dfrac{A}{(N+S)} \le l_{p_i} \le (l_u+l_g+1)$ and $\lambda \le l_{p_i}< l_u$. In mFHMRS, $p_1, p_2,...,p_{N+S}$ and $u$ are the secret parameters. $n$ is also a secret. The size of secret keys to prevent possible attacks and ensure the correctness of decryption are given as follows.\\
($l_p=l_{p_i}$: number of bits in $p_i$, $l_u$: number of bits in $u$, $l_g$: number of bits in $g_i$, random number, $l_m$: number of bits in input message $m$, $l_{\mathbb{M}}$: number of bits in message space $\mathbb{M}$, )
\begin{itemize}
\item $\dfrac{(N+1)}{(N+S)}*(l_u +l_g+1)+\dfrac{A}{(N+S)} \le l_{p_i} \le (l_u+l_g+1)$ for the correctness of decryption after homomorphic operations (Ref. \ref{gcod}) and to resist attack specified in section \ref{gkpa}. Choose $S$ such that $\dfrac{(N+1)}{(N+S)}*(l_u +l_g+1)+\dfrac{A}{(N+S)} \le l_{p_i} \le (l_u+l_g+1)$ and $\lambda \le l_{p_i}< l_u$ to prevent attack mentioned in section \ref{gsle} and \ref{gslp}. 
\item  $l_{p_i} \ge \lambda$ to prevent guessing $p_i$ value from the ciphertext as $i^{th}$ element of ciphertext $\in$ $(0, 1, \dots, p_{i}-1)$ (Ref. \ref{grg}).  
\item  $l_{p_i} \le (l_u+l_g+1)$ to mitigate the lattice attack mentioned in  \ref{coa-lba} and \ref{kpa-lba}.\\
\item $l_{u} > (l_{\mathbb{M}}=(N+1)* l_m + A)$ for the correctness of decryption as given in section \ref{gcod}.
\item $l_g \ge \lambda/4$ and $l_u>l_p $ to prevent attack based on solving linear equations as given in sections \ref{gsle} and \ref{gslp}.\\
\end{itemize}
For instance,  If $\lambda = 128, l_m= 10, N=1$ and $A= 20$, the size of secret parameters can be $l_{u}= 130$, $l_{g}= 42$, $l_{p}= 128$, $S$=2 to satisfy the above criteria including $122 \le l_{p_i} \le 173$. If $\lambda = 128, l_m=10, N=14$ and $A= 30$,  the size of secret parameters can be $l_{u}= 182$, $l_{g}= 42$,$l_{p}= 180$ and  $S=5$ to satisfy the above criteria including $179.2 \le l_{p_i} \le 225$.
\item \textbf{Encrypt ($m_{i}, p_1, p_2,...,p_{N+S}, u$):} Accepts secret primes ($p_1, p_2,...,p_{N+S}$ and $u$) and the $i^{th}$ message $m_{i}$ $\in$ $\mathbb{Z}_{\mathbb{M}}$ to be encrypted as inputs. Generate random number $g_{i}$ of length $l_g$ bits using a cryptographically secure pseudo-random number generators (CSPRNG). The ciphertext $c_{i} = (c_{i1}, c_{i2},..., c_{i(N+S)})$ is computed relying on Chinese Remainder Theorem (CRT) as shares of $m_{i}$ as follows.
\begin{equation}
\label{eqn:e4}
\begin{split}
c_{i}=& (c_{i1}, c_{i2},..., c_{i(N+S)})\\=&([(m_{i}+(g_{i}* u))]_{p_1}, [(m_{i}+(g_{i}* u)]_{p_2}, ..., \\&[(m_{i}+(g_{i}* u)]_{p_{N+S}})
\end{split}
\end{equation}
where $[]_{p_i}$ indicates modulo $p_i$ operation. $c_{i1}$  represents $[(m_{i}+(g_{i}* u))]_{p_1}$.\\
For instance,  assume mFHMRS supports single multiplication, then $S=2$  and $p_1, p_2$ and $p_{3}$ are 128-bit primes,  ciphertext size will be a maximum of  384 bits since ciphertext size depends on the size of $p_1, p_2$ and $p_{3}$. If mFHMRS supports 14 multiplications, $l_{p_i}=180$ and S=5, ciphertext size will be a maximum of 3420 bits.
\item \textbf{Decrypt ($c_{i}, p_1, p_2,...,p_{N+S}, u$) :} This function takes the ciphertext $c_{i} = (c_{i1}, c_{i2},..., c_{i(N+S)})$ corresponding to $i^{th}$ message $m_{i}$ and secret primes ($ p_1, p_2,...,p_{N+S}$ and $u$) of mFHMRS as inputs. Based on CRT, the message ($m_{i}$) is computed from $c_{i}$ as in Eqn. \ref{eqn:e5}.
\begin{equation}
\label{eqn:e5}
\begin{split}
m_{i}&=\big[ \big[([c_{i1}]_{p_1} * p'_1* [p'^{-1}_{1}]_{p_1}) +( [c_{i2}]_{p_2}* p'_2* [p'^{-1}_{2}]_{p_2})\\&,...,\\& ( [c_{i(N+S)}]_{p_{(N+S)}}* p'_{(N+S)}* [p'^{-1}_{(N+S)}]_{p_{(N+S)}})\big ]_{ n}\big]_{u}\\&=\big[ CRT( [c_{i1}]_{p_1}, [c_{i2}]_{p_2},...,[c_{i(N+S)}]_{p_{(N+S)}}) \big]_{u}
\end{split}
\end{equation}
Where, $ p'_1= p_2 * p_3*...* p_{(N+S)}$,  $ p'_2= p_1 * p_3*...* p_{(N+S)}$  and $ p'_{(N+S)}= p_1 * p_2*...* p_{(N+S-1)}$\\
\end{itemize}
\subsection{Homomorphic Properties}
mFHMRS supports homomorphic addition, homomorphic constant addition, homomorphic constant multiplication and homomorphic multiplication. In FHMRS, homomorphic operations do not involve modular operation. For the correctness of decryption on the homomorphically operated ciphertext, the modular operation is performed in the decryption function before performing CRT-based reconstruction. Let $c_{1} = (c_{11}, c_{12},..., c_{1(N+S)})$ and $c_{2} = (c_{21}, c_{22},..., c_{2(N+S)})$ be the mFHMRS encryption of $m_1$ and $m_2$, respectively. Homomorphic properties of mFHMRS are given below.
\begin{itemize}
\item Homomorphic addition: mFHMRS encryption of $m_1$+ $m_2$ can be computed as in Eqn. (\ref{eqn:gea})
\begin{equation}
\label{eqn:gea}
\begin{split}
&c_{add}= \big(c_{11}+ c_{21}, c_{12}+c_{22},..., c_{1(N+S)}+c_{2(N+S)}\big)
\end{split}
\end{equation}
\item Homomorphic multiplication: mFHMRS encryption of $m_1$ $*$ $m_2$ can be computed as in Eqn. (\ref{eqn:gem})
\begin{equation}
\label{eqn:gem}
\begin{split}
&c_{mul}= \big(c_{11}* c_{21}, c_{12}* c_{22},..., c_{1(N+S)}* c_{2(N+S)}\big)
\end{split}
\end{equation}
\item Homomorphic constant addition: mFHMRS encryption of $a+ m_1$, where $a$ is a scalar, can be computed as follows.
\begin{equation}
\label{eqn:gsa}
\begin{split}
&c_{const-add}= \big(c_{11}+ a, c_{12}+a,..., c_{1(N+S)}+a\big)
\end{split}
\end{equation}
\item Homomorphic constant multiplication: mFHMRS encryption of $a* m_1$ can be computed as follows.
\begin{equation}
\label{eqn:gsm}
\begin{split}
&c_{const-mul}= \big(c_{11}* a, c_{12}* a,..., c_{1(N+S)}* a\big)
\end{split}
\end{equation}
\end{itemize}
$+_{a} (-_{ a} )$ represents homomorphic addition (subtraction) of mFHMRS scheme.
$+_{ca} (-_{ca})$
represents homomorphic constant addition (subtraction) in mFHMRS. $*_{m}$ represents the multiplicative homomorphism of the mFHMRS scheme. $*_{cm}$
represents homomorphic constant multiplication in the mFHMRS scheme.
\section{Security analysis of modified FHMRS (mFHMRS)}
The security of the proposed mFHMRS is analyzed based on the computational difficulty of recovering the message ($m_{i}$) from the corresponding ciphertext tuple $c_{i} = (c_{i1}, c_{i2},..., c_{i(N+S)})$. Retrieving  $m_{i}$ from this tuple requires knowing the secret primes ($ p_1, p_2,...,p_{N+S}$ and $u$). Consequently, the security of the mFHMRS scheme (protecting these secret prime values) against brute-force attacks, lattice-based attacks and linear equation solving is analysed. The number of prime numbers less than $2^l$ is approximately, $ \dfrac{2^{l} - 1}{ ln(2^{l} - 1)}$ \cite{prime}. Therefore, the approximate number of primes with exactly $l$ bits is given by, $ d_{l} = \dfrac{2^{l} - 1}{ ln(2^{l} - 1)}-\dfrac{2^{l-1} - 1}{ ln(2^{l-1} - 1)}$. Let the bit-lengths of $p_i$ in $ p_1, p_2,...,p_{N+S}$ be $l_{p}$-bits and the bit-length of $u$ be $l_{u}$-bits. $d_{l_{p}}$ and $d_{l_{u}}$ represent the number of prime numbers of bit-length $l_{p}$ ( for $p_i$ in $ p_1, p_2,...,p_{N+S}$) and $l_{u}$ (for $u$), respectively. The following sections analyse the resistance of the mFHMRS scheme against different possible attacks.
\subsection{Ciphertext-only Attacks}
In this attack, the adversary obtains the $N+S$ CRT shares of the messages ($c_{i} = (c_{i1}, c_{i2},..., c_{i(N+S)}))$.
\subsubsection{Brute-force Attack} Given the known ciphertexts, the attacker attempts all possible combinations of the secret keys $ p_1, p_2,...,p_{N+S}$, and $u$ until the correct set of secrets is found. The strength of the encryption scheme against brute-force attacks is determined by the size of the effective keyspace, which is defined by the number of possible values for $ p_1, p_2,...,p_{N+S}$ and $u$. Accordingly, the total number of brute-force trials required to uncover the secrets is $(d_{l_{p}})^{(N+2)} *d_{l_{u}}$, where $d_{l_{p}} \approx \dfrac{2^{l_{p}} - 1}{ ln(2^{l_{p}} - 1)}-\dfrac{2^{l_{p}-1} - 1}{ ln(2^{l_{p}-1} - 1)}$ and $ d_{l_{u}} \approx \dfrac{2^{l_{u}} - 1}{ ln(2^{l_{u}} - 1)}-\dfrac{2^{l_{u}-1} - 1}{ ln(2^{l_{u}-1} - 1)}$.
For instance,  assume mFHMRS supports single multiplication, $p_1, p_2, p_{3}$, and $u$ are the secret primes. If $p_1, p_2$ and $p_{3}$ are 128-bit primes and $u$ is a 130-bit prime, the brute-force attack complexity reaches approximately $ 2^{490}$.
\subsubsection{Lattice-based Attack}
\label{coa-lba}
In this section, we consider the lattice-based attack described in \cite{gdhgv, gacd}. In this attack, an attacker constructs a lattice based on the available mFHMRS ciphertexts and tries to get a short vector which reveals the secret parameters of the mFHMRS scheme. Here, we can rewrite $c_{i1}=[(m_{i}+(g_{i}* u))]_{p_1}$ in \ref{eqn:e4} as $c_{i1}=m_{i}+g_{i}* u- k_{i1}*p_1$, where $u$ and $p_1$ are part of secret keys of mFHMRS and $k_{i1}=\lfloor \dfrac{m_{i}+g_{i}* u}{p_1} \rfloor$. We can consider finding $u$ from $c_{i1}=m_{i}+g_{i}* u- k_{i1}*p_1$ as approximate common divisor problem (ACDP) by considering $m_{i}-k_{i1}*p_1$ as error term \cite {gdhgv, gacd}.  Let the error term $m_{i}-k_{i1}*p_1$ is denoted as $r_i$. Then $c_{i1}$ can be written as in Eqn. (\ref{gll}).
\begin{equation}
\label{gll}
c_{i1}=g_{i}* u+r_{i1}
\end{equation}
Attacker knows only $c_{i1}$ and trying to approximate $\dfrac{g_{i}}{g_0}$, where $g_0$ is the random number associated with  $c_{01}=g_{0}* u+r_{01} $. I.e., the fractions $\dfrac{g_{i}}{g_0}$ are an instance of simultaneous Diophantine approximation to $\dfrac{c_{i}}{c_0}$. This attack does not benefit significantly from having an exact sample $c_0 = u*g_0$. Hence, we consider $c_0 = u*g_0+r_{01}$.
If the attacker can find $g_0$ then he can solve the ACDP by computing $r_{01} \equiv c_{01} \bmod g_0$ and hence $(c_{01} - r_{01})/g_0 = u$ \cite{gacd}.
Similar to \cite{gdhgv, gacd}, we can consider the LLL algorithm for finding the short vectors and construct (t+1)-dimensional Lattice $L$ spanned by the rows of basis matrix $\textbf{B}$.\\
\begin{align*}
\textbf{B} = \begin{bmatrix}
2^{l_k+l_{p_1}} & c_{11} &c_{21}&\dots&c_{t1}\\
& -c_{01} &&&\\
&& -c_{01}&&\\
& &&\ddots&\\
& & &&-c_{01}\\
\end{bmatrix}
\end{align*}
$L$ contains the vector, $v= (g_0, g_1, \dots, g_t) \textbf{B}$.

\begin{align*}
\begin{split}
v&=g_0*(2^{l_k+l_{p_1}} , c_{11} ,c_{21},\dots,c_{t1})+g_1*(0,-c_{01},\dots,0 )\\&+ \dots+g_t*(0, 0, \dots, -c_{01})\\
&=(g_0*2^{l_k+l_{p_1}} , g_0*c_{11}-g_1*c_{01} ,\dots, g_0*c_{t1}-g_t*-c_{01})\\
&=(g_0*2^{l_k+l_{p_1}} , g_0*r_{11}-g_1*r_{01} ,\dots, g_0*r_{t1}-g_t*-r_{01})\\
\end{split}
\end{align*}
The length (Euclidean norm) of v' is $\sqrt{t+1}*2^{l_{g_{i}}+ l_{k_{i1}}+l_{p_1}}$, where $l_{k_{i1}}\approx ( l_g+l_u-l_p)$.
LLL algorithm gives shortest vectors of $L$ which are lesser than $\sqrt{\dfrac{t+1}{2\pi e}}*(det(L))^{1/(t+1)}$. The determinant of the upper triangular matrix L, Det(L), is approximately given by $2^{l_k+l_{p_1}}*(2^{l_{p_1}})^{t}$$\approx $$2^{((t+1)*l_{p_1})+l_k}$, here maximum value of $c_{10}$ is taken as approximately $2^{l_{p_1}}$ and $l_{k}\approx ( l_g+l_u-l_p)$.
Then the reduced basis vectors based on LLL will have a length lesser than $\sqrt{\dfrac{t+1}{2\pi e}}*2^{l_{p_1}+(l_k/(t+1))}$. The selected length of $p_1$ and $g_i$ makes length of $v$ (Euclidean norm of v=$\sqrt{t+1}*2^{l_{g_{i}}+ l_{k_{i1}}+l_{p_1}}$) greater than $\sqrt{\dfrac{t+1}{2\pi e}}*(det(L))^{1/(t+1)}$=$\sqrt{\dfrac{t+1}{2\pi e}}*2^{l_{p_1}+(l_k/(t+1))}$. Hence, an attacker cannot get $g_0$ based on the lattice basis reduction method. Therefore, mFHMRS resist finding $u$ based on LLL algorithm.\\
Similarly,  we can construct a lattice by considering the error term, $r_i$, as $m_{i}+g_{i}* u$ and  $c_{i1}=-k_{i1}*p_1+r_{i1}$ to find $p_1$. 
We can consider the LLL algorithm for finding the short vectors and construct (t+1)-dimensional Lattice $L'$ spanned by the rows of basis matrix $\textbf{B'}$.\\
\begin{align*}
\textbf{B'} = \begin{bmatrix}
2^{l_u+l_{g}} & c_{11} &c_{21}&\dots&c_{t1}\\
& -c_{01} &&&\\
&& -c_{01}&&\\
& &&\ddots&\\
& & &&-c_{01}\\
\end{bmatrix}
\end{align*}
$L'$ contains the vector, $v'= (k_{01}, k_{11}, \dots, k_{t1}) \textbf{B'}$.

\begin{align*}
\begin{split}
v'&=k_{01}*(2^{l_u+l_{g}} , c_{11} ,c_{21},\dots,c_{t1})+k_{11}*(0,-c_{01},\dots,0 )\\&+ \dots+k_{t1}*(0, 0, \dots, -c_{01})\\
&=(k_{01}*2^{l_u+l_{g}} , k_{01}*c_{11}-k_{11}*c_{01} ,\dots, k_{01}*c_{t1}-k_{t1}*-c_{01})\\
&=(k_{01}*2^{l_u+l_{g}} , k_{01}*r_{11}-k_{11}*r_{01} ,\dots, k_{01}*r_{t1}-k_{t1}*-r_{01})\\
\end{split}
\end{align*}
The length (Euclidean norm) of v is $\sqrt{t+1}*2^{l_{g_{i}}+ l_{k_{i1}}+l_{u}}$, where $l_{k_{i1}}\approx  l_g+l_u-l_p$.
LLL algorithm gives shortest vectors of $L$ which are lesser than $\sqrt{\dfrac{t+1}{2\pi e}}*(det(L'))^{1/(t+1)}$. The determinant of the upper triangular matrix L', Det(L'), is approximately given by $2^{l_u+l_{g}}*(2^{l_{p_1}})^{t}$, here maximum value of $c_{10}$ is taken as approximately $2^{l_{p_1}}$ and $l_{k}\approx ( l_g+l_u-l_p)$.
Then the reduced basis vectors based on LLL will have a length lesser than $\sqrt{\dfrac{t+1}{2\pi e}}*2^{(t*l_{p_1}/t+1)+(l_u+l_{g}/(t+1))}$. Since, $l_u+l_{g} >l_{p_1}$, length of $v'$ (Euclidean norm of v'=$\sqrt{t+1}*2^{l_{g}+l_{u}+ l_g+l_u-l_p}$) greater than $\sqrt{\dfrac{t+1}{2\pi e}}*(det(L))^{1/(t+1)}$=$\sqrt{\dfrac{t+1}{2\pi e}}*2^{(t*l_{p_1}/t+1)+(l_u+l_{g}/(t+1))}$. Hence, LLL will not help to find $k_{01}$ and hence the attacker will not get $p_1$.  Similarly, based on the lattice basis reduction method, an attacker cannot solve for $ p_2,...,p_{N+S}$. Hence, mFHMRS is resistant to lattice basis reduction-based attacks.
\subsection{Known-plaintext Attacks}
If an attacker-such as an eavesdropper monitoring communications via the gateway or the cloud-has access to ciphertext-plaintext pairs, they may launch a Known Plaintext Attack (KPA) to recover the secret keys. The KPA against FHMRS given in section \ref{gkpa} could not be applied on mFHMRS since ciphertext will not reveal $(m_{i}+(g_{i}* u))$ due to the modification. In the context of the mFHMRS scheme, a KPA involves the attacker attempting to deduce the secret values  $ p_1, p_2,...,p_{N+S}$ and $u$ using known values of  $m_i$ and the corresponding ciphertext ($c_{i} = (c_{i1}, c_{i2},..., c_{i(N+S)}))$.
\subsubsection{Brute-force Attack}
In this KPA against mFHMRS, an attacker can exhaustively solve Eqn. (\ref{eqn:e5}) across all possible combinations of  $p_1, p_2,...,p_{N+S}$ and $u$ by aiming to extract these secrets from the known pair  $m_{i}$ and ($c_{i} = (c_{i1}, c_{i2},..., c_{i(N+S)}))$. The steps involved in carrying out this KPA are outlined as follows
\begin{enumerate}
\item Select a ciphertext ($c_{i} = (c_{i1}, c_{i2},..., c_{i(N+S)}))$ and  its corresponding message $m_{i}$.
\item Predict one set of secret keys as $ p_1, p_2,...,p_{N+S}$ and $u$.
\item Put the predicted values in step 2 in the Eqn. (\ref{eqn:e5}) and check whether Eqn. (\ref{eqn:e5}) holds correctly.
\item Continue step 3 for all possible values of $ p_1, p_2,...,p_{N+S}$ and $u$ until the equation's right side matches the message $m_{i}$.
\item The values of prime numbers $ p_1, p_2,...,p_{N+S}$ and $u$ that satisfy the Eqn. (\ref{eqn:e5}) are the secret parameters of the mFHMRS scheme.
\end{enumerate}
Since the minimum and maximum number of trails to obtain secret parameters $ p_1, p_2,...,p_{N+S}$ and $u$ is one and $(d_{l_{p}})^{(N+S)} *d_{l_{u}}$, respectively, on an average, attackers have to try $((d_{l_{p}})^{(N+S)} *d_{l_{u}}+ 1)\big/2$ possible values.  Hence, the total key space for this KPA is approximately equal to $((d_{l_{p}})^{(N+S)} *d_{l_{u}})\big/2$.
For instance,  assume mFHMRS supports single multiplication, $p_1, p_2, p_{3}$, and $u$ are the secret primes.
If $p_1, p_2$ and $p_{3}$ are 128-bit primes and $u$ is a 130-bit prime, the computational complexity of this KPA will be approximately $ 2^{489}$.\\
\subsubsection{Lattice-based Attack}
\label{kpa-lba}
As given in \ref{coa-lba},  $c_{i1}=m_{i}+g_{i}* u- k_{i1}*p_1$. Since attacker knows $c_{i1}$ and corresponding $m_{i}$, a lattice $L_1$ can be formed based on the basis matrix $\textbf{B}_1$ with known $c_{i1}- m_{i}$, where $c_{i1}-m_{i}=g_{i}* u- k_{i1}*p_1$. Here error term,  is  $r'_{i1}=$ $-k_{i1}*p_1$.
\begin{align*}
\textbf{B}_1=\begin{bmatrix}
2^{l_k+l_{p_1}} & c_{11}- m_{1}&\dots&c_{1t}-m_{t}\\
& -(c_{10}-m_{0}) &&\\
& &\ddots&\\
& & &-(c_{10}-m_{0})\\
\end{bmatrix}
\end{align*}
$L_1$ contains the vector, $v'= (g_0, g_1, \dots, g_t) \textbf{B}_1=(g_0*2^{l_k+l_{p_1}} , g_0*r'_{11}-g_1*r'_{10} ,\dots, g_0*r'_{1t}-g_t*-r'_{10})$. Since the Euclidean norm of $v'$ is approximately equal to that of $v$ and the determinant of $L$ is approximately equal to that of $L_1$, LLL-based lattice basis reduction will not reveal $u$. Similar to analysis in \ref{coa-lba}, in this KPA, the LLL-based method will not reveal the secrets $ p_1, p_2,...,p_{N+S}$ and $u$.
\subsubsection{Attack based on  solving linear equations in u and $p_i$s}
\label{gsle}
When an attacker has ciphertexts $ (c_{11}, c_{12},..., c_{1(N+S)})$ and $ (c_{21}, c_{22},..., c_{2(N+S)})$ corresponding to $m_1$ and $m_2$, the following attack can be done to retrieve the secrets $ p_1, p_2,...,p_{N+S}$ and $u$. Since $c_{ij}-m_{i}=g_{i}* u- k_{ij}*p_j$, where $1\le i\le 2$ and $1\le j\le N+S$. The steps involved in carrying out this KPA are given as follows
\begin{enumerate}
\item Select two ciphertexts $ (c_{11}, c_{12},..., c_{1(N+S)})$ and $ (c_{21}, c_{22},..., c_{2(N+S)})$ and its corresponding plaintexts $m_1$ and $m_2$
\item An attacker can solve the linear equations in Eqn. (\ref{eqn:ge4}) for different values of $g_1, g_2, k_{2i}, k_{1i}$. E.g., Predict one set of values for $g_1, g_2, k_{21}, k_{11}$ and solve equations in Eqn. (\ref{eqn:ge4}) to get a set of  secrets $ p_1$ and $u$.  Predict one set of values for $g_1, g_2, k_{22}, k_{12}$ and solve equations in Eqn. (\ref{eqn:ge4}) to get a set of  secrets $ p_2$ and $u$\\
\begin{equation}
\label{eqn:ge4}
\begin{split}
&c_{1i}-m_{1}=g_{1}* u- k_{1i}*p_i\\
&c_{2i}-m_{2}=g_{2}* u- k_{2i}*p_i\\
\end{split}
\end{equation}
\item Continue steps 2  for all possible values of $g_1, g_2, k_{2i}, k_{1i}$. The values of prime numbers $ p_1, p_2,...,p_{N+S}$ that corresponding to the same $u$ will satisfy the Eqn. (\ref{eqn:e5}) and that are the actual values of the secret parameters $p_1, p_2,...,p_{N+S}$ and $u$ of the mFHMRS scheme.
\end{enumerate}
Since the possibilities of $g_i$ are $2^{l_g}$ and possibilities of $k_{ij}$ are $ l_g+l_u-l_p$, maximum number of trails to obtain secret parameters $ p_1, p_2,...,p_{N+S}$ and $u$ is $2^{l_g}*2^{l_g}*(2^{(2)*( l_g+l_u-l_p)}$=$2^{4*l_g+2*(l_u-l_p)}$.
The minimum number of trails to obtain secret parameters $ p_1, p_2,...,p_{N+S}$ and $u$ is one. Hence, on average, attackers have to try $2^{4*l_g+2*(l_u-l_p)}/2$ possible values. Hence, the total key space for KPA is approximately equal to $2^{4*l_g+2*(l_u-l_p)-1}$. Hence, the total trails for this KPA is greater than $(2^{4*l_g})$. Hence, this attack can be mitigated if we choose $l_g \ge \lambda/4$\\

There can be more stringent constraint on the selection of bit length of $g_i$, if the $g_i$ s are very close. When an attacker has ciphertexts $ (c_{11}, c_{12},..., c_{1(N+S)})$ $ (c_{21}, c_{22},..., c_{2(N+S)})$, $ (c_{31}, c_{32},..., c_{3(N+2)})$ and $ (c_{41}, c_{42},..., c_{4(N+2)})$ corresponding to $m_1=0$, $m_2=0$, $m_3=0$ and $m_4=0$, the following attack can be done to retrieve the secrets $ p_1, p_2,...,p_{N+S}$ and $u$. 
\begin{equation}
\label{eqn:ge5}
\begin{split}
c_{1i}-c_{2i}=&(g_{1}-g_2)* u- (k_{1i}-k_{2i})*p_i\\
c_{3i}-c_{4i}=&(g_{3}-g_4)* u- (k_{3i}-k_{4i})*p_i\\
\end{split}
\end{equation}
Though the possibilities of $k_{ij}$ are $(2^{(l_g+l_u-l_p)}$, the possibilities of $k_{1j}-k_{2j}$ varies based on $|g_{1}-g_2|$. Also, for instance, if $|g_{1}-g_2|$$<p_1/u$, $k_{11}=k_{21}$ (proof:$k_{11}=k_{21}$ $\implies  \lfloor \dfrac{g_{1}* u}{p_1} \rfloor$ $=\lfloor \dfrac{g_{2}* u}{p_1} \rfloor$$\implies -1 < \dfrac{g_{1}* u}{p_1} $ $- \dfrac{g_{2}* u}{p_1} <1$$\implies|g_{1}-g_2|$$<p_1/u$). If $|g_{1}-g_2|$$<p_1/u$, $k_{11}=k_{21}$ and GCD of $(c_{11}-c_{21}, c_{31}-c_{41})$ (refer Eqn. \ref{eqn:ge5}) reveals $u$. Hence, $ l_u> l_{p_{i}}$ to avoid revealing $u$. Since $g_{i}$ values are integers,  $|g_{1}-g_2|$$>p_1/u$ and $k_{11}\ne k_{21}$. Since random numbers $g_{i}$s are generated using Cryptographically secure pseudo-random number generators (CSPRNG), $ (g_{1}-g_2)$ is unpredictable and can range from any value between one and $2^{l_g }-1$ since $g_{i}$ are $l_g$ bits.

I.e.,  if  $ u >p_{i}$, for different $g_i $,  $k_i$ will be different. As difference in $g_i $ is unpredictable, difference in $k_i$ will also be unpredictable. Also, as $g_i$ varies $k_i$ also varies. Since $g_{i}$s are generated using CSPRNG, the possibilities of $| (g_{1}-g_2)|$ are similar to possibilities of $g_i $ (i.e., $2^{l_g} -1$) and  the possibilities of $| (k_{1i}-k_{2i})|$ are similar to possibilities of $k_i$ (i.e., $2^{(l_g+l_u-l_p)}$).  Hence, to solve $u$ and $p_{i}$ from Eqn. \ref{eqn:ge5}, the attacker has to try $2^{4 *l_g +2*(l_u-l_p)-1}$ possibilities on average.
\subsubsection{Attack based on  solving linear equations in  $p_i$s}
\label{gslp}
Let $(c_{11}, c_{12},..., c_{1(N+S)})$ and $(c_{21}, c_{22},..., c_{2(N+S)})$ be the mFHMRS encryption of $m_1$ and $m_2$, respectively. Then  $c_{11}$ and $c_{12}$, the  first two elements of ciphertext tuple corresponding to $m_1$ can be written as in Eqn. (\ref{eqn:ge8}).
\begin{equation}
\label{eqn:ge8}
\begin{split}
&c_{11}=m_{1}+g_{1}* u- k_{11}*p_1\\
&c_{12}=m_{1}+g_{1}* u- k_{12}*p_2\\
\end{split}
\end{equation}
The difference $(c_{11}-c_{12})$ eliminates the common term  $m_{1}-g_{1}* u$ and gives a relation between $p_1$ and $p_2$ as given in Eqn. (\ref{eqn:ge9}). 
\begin{equation}
\label{eqn:ge9}
\begin{split}
c_{11}-c_{12}=&- k_{11}*p_1+k_{12}*p_2\\
\end{split}
\end{equation}
Similarly, with the ciphertext components $c_{21}$ and $c_{22}$ corresponding to $m_2$ , we get  $c_{21}-c_{22}=-k_{21}*p_1+k_{22}*p_2$. By computing $(c_{11}-c_{12})-(c_{21}-c_{22})$, we will get first equation in Eqn. (\ref{eqn:ge10}). 
Similarly, with the ciphertext components corresponding to $m_3$ and $m_4$, we get second equation in Eqn. (\ref{eqn:ge10}). 
\begin{equation}
\label{eqn:ge10}
\begin{split}
((c_{11}-c_{12})-(c_{21}-c_{22}))=&- (k_{11}-k_{21})*p_1\\&+(k_{12}-k_{22})*p_2\\
((c_{31}-c_{32})-(c_{41}-c_{42}))=&- (k_{31}-k_{41})*p_1\\&+(k_{32}-k_{42})*p_2\\
\end{split}
\end{equation}
Since, $k_{i1}=\lfloor \dfrac{m_{i}+g_{i}* u}{p_1} \rfloor$ and $ u >p_{i}$, differences in $g_i$ causes differences in $k_i$.
As $g_i $ is random, $k_i$ is also random and hard to predict. The possibilities of $| (k_{1i}-k_{2i})|$ are similar to possibilities of $k_i$ (i.e., $2^{(l_g+l_u-l_p)}$). Hence for solving $p_1$ and $p_2$, attacker has to try $2^{4 *(l_g+l_u-l_p)-1}$ possibilities on average.

\subsection{Resistance to guessing $p_i$ values from $c_{i}$}
\label{grg}
The possibility of guessing $p_i$, where $i$ ranges from $1$ to $(N+S)$, from the corresponding ciphertext share (e.g., $p_1$ from $c_{i1}$) can be mitigated by selecting large $p_i$ and changing the secret primes so that attacker could not collect enough ciphertexts. Here, we have considered $p_i$ s are of length $\ge \lambda$ to resist this attack.
\subsection{Lattice -based recovery of integer from noisy Chinese remaindering with known $p_i$s}
In \cite{gncr}, the following two problems related to Chinese remaindering are considered.\\
\textbf{Problem 1 (Noisy Chinese remaindering):} Let $0 \le X \le B$, and $p_1, \dots, p_n$ be
coprime integers. Given $n$ sets $S_1, \dots, S_n$ where each $S_i = \{r_{i,j}\}_{1\le j \le m}$ contains
$m - 1$ random elements in $ \mathbb{Z}_{p_i}$ and $X \bmod p_i$, recover the integer X, provided
that the solution is unique.\\
\textbf{Problem 2 (Chinese remaindering with errors):} Given as input integers $t$, $B$
and $n$ points $(r_1, p_1),\dots,  (r_n, p_n) \in \mathbb{N}^2$ where the $p_i$'s are coprime, output all
numbers $0 \le X < B$ such that $X\cong r_i (\bmod p_i)$ for at least $t$ values of $i$.\\

These problems consider $p_i$s are known, and exact $X \bmod p_i$ s are unknown and recover $X$. Lattice reduction method is applied on the Lattice developed based on the known $p_i$s to recover the integer X.\\

However, in mFHMRS, $X \bmod p_i$ s are known, and $p_i$s are unknown. Hence, the above-mentioned lattice-based integer recovery is not relevant to mFHMRS. Hence, we have considered a lattice attack, where a lattice is developed based on the known ciphertext, which is similar to $X \bmod p_i$ to recover the secrets as given in Section \ref{coa-lba} and \ref{kpa-lba}.
\section{Correctness of Decryption}
\label{gcod}
For the correctness of decryption after $N$ consecutive multiplication, CRT based reconstruction should return $\big(m_{1}+ (g_{1}*u) \big) * \big(m_{2}+ (g_{2}*u) \big)*\dots *\big(m_{N+1}+ (g_{N+1}*u)\big)$ so that modular reduction by $u$ gives $m_{1}*m_{2}*\dots*m_{N+1}$. Hence,  $\big(m_{1}+ (g_{1}*u) \big) * \big(m_{2}+ (g_{2}*u) \big)*\dots *\big(m_{N+1}+ (g_{N+1}*u)\big) < n$ and $(N+1)*(l_g+l_u+1) < l_n$. If there are A consecutive additions to be supported, then $((N+1)*(l_g+l_u+1)) + A < l_n$. Since $p_1* p_2*\dots*p_{N+S}=n$ and $p_1, p_2,...,p_{N+S}$ are equal in size, bit length of $p_i$, $l_{p_{i}}>\dfrac{((N+1)*(l_g+l_u+1)) +A} {N+S}$.  \\
Also, after CRT-based reconstruction, modular reduction by $u$ is performed to retrieve the message. Hence, after homomorphic operations, the message size should be less than $u$ for the correctness of decryption. I.e.,  If $l_m$ is the size of input messages and there are N consecutive multiplications and A consecutive additions,  $l_{\mathbb{M}}=log_{2} (\mathbb{M}) > (N+1)* l_m + A$ and $l_u > l_{\mathbb{M}} $, where $\mathbb{M}$ is the message space.

\end{document}